\documentclass[Physsubmission, Phys]{SciPost}

\hypersetup{
    colorlinks,
    linkcolor={red!50!black},
    citecolor={blue!50!black},
    urlcolor={blue!80!black}
}

\usepackage{microtype}
\usepackage[bitstream-charter]{mathdesign}
\DeclareSymbolFont{usualmathcal}{OMS}{cmsy}{m}{n}
\DeclareSymbolFontAlphabet{\mathcal}{usualmathcal}

\begin{document}

\begin{center}{\Large \textbf{
Photoproduction of diffractive dijets in \textsc{Pythia~8}\\
}}\end{center}

\begin{center}
Ilkka Helenius\textsuperscript{*,1,2}
\end{center}

\begin{center}
{\bf 1} University of Jyvaskyla, Department of Physics, P.O. Box 35, FI-40014 University of Jyvaskyla, Finland \\
{\bf 2} Helsinki Institute of Physics, P.O. Box 64, FI-00014 University of Helsinki, Finland \\
* ilkka.m.helenius@jyu.fi
\end{center}

\begin{center}
\today
\end{center}


\definecolor{palegray}{gray}{0.95}
\begin{center}
\colorbox{palegray}{
  \begin{tabular}{rr}
  \begin{minipage}{0.1\textwidth}
    \includegraphics[width=22mm]{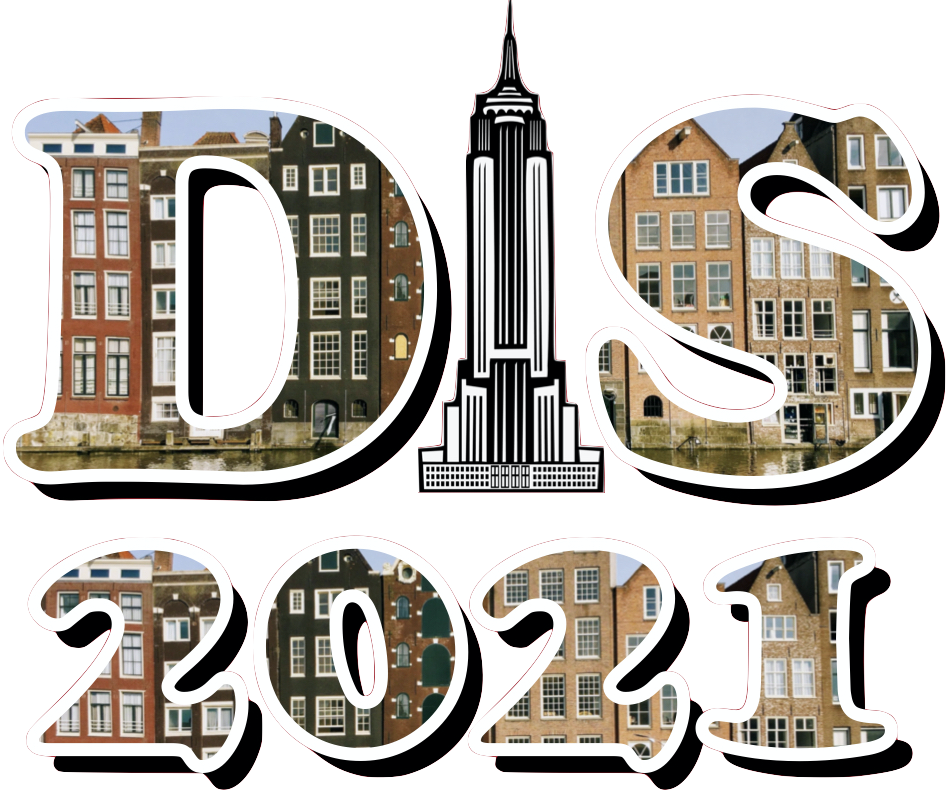}
  \end{minipage}
  &
  \begin{minipage}{0.75\textwidth}
    \begin{center}
    {\it Proceedings for the XXVIII International Workshop\\ on Deep-Inelastic Scattering and
Related Subjects,}\\
    {\it Stony Brook University, New York, USA, 12-16 April 2021} \\
    \doi{10.21468/SciPostPhysProc.?}\\
    \end{center}
  \end{minipage}
\end{tabular}
}
\end{center}

\section*{Abstract}
{\bf
  The new framework for the simulations of hard diffractive events in photoproduction within \textsc{Pythia}~8 is presented. The model, originally introduced for proton-proton collisions, applies the dynamical rapidity gap survival probability based on the multiparton interaction model in \textsc{Pythia}. These additional interactions provide a natural explanation for the observed factorization-breaking effects in hard diffraction by filling up the rapidity gaps used to classify the events of diffractive origin. The generated cross sections are well in line with the existing data from HERA experiments and predictions for the future electron-ion collider are presented. In addition, also predictions for ultra-peripheral collisions at the LHC have been calculated where the factorization-breaking effects are found to be more pronounced.
}

\section{Introduction}
In the Ingelman-Schlein approach \cite{Ingelman:1984ns} for hard diffraction the diffractive cross sections can be factorized into diffractive PDFs and perturbatively calculable partonic coefficient functions. The diffractive PDFs can be determined in a global QCD analysis using data for diffractive processes in DIS in a similar manner as for inclusive PDFs. However, it has been observed that this factorization breaks down in proton-proton collisions as the predicted cross sections overshoot the data by an order of magnitude \cite{CDF:2000rua, ATLAS:2015yqo, CMS:2020dns}. Similarly, factorization-based calculations for diffractive dijets in photoproduction at HERA tend to lie a factor of two above the H1 and ZEUS measurements, depending on the considered kinematics \cite{H1:1998bvm, H1:2007jtx, ZEUS:2007uvk, H1:2015okx}. In this work we study such factorization-breaking effects within \textsc{Pythia}~8 \cite{Sjostrand:2014zea} Monte Carlo event generator.

In the recently introduced \textsc{Pythia}~8 model for hard diffraction the factorization breaking effects are modelled with the dynamically calculated rapidity gap survival probabilities \cite{Rasmussen:2015qgr, Helenius:2019gbd}. Such rapidity gaps are typically used as an experimental signature for diffractive events where the gap is a consequence of a colour-neutral Pomeron exchange. Following the idea by Bjorken \cite{Bjorken:1992er}, such rapidity gaps may be filled by the particles from additional partonic interactions in the collision event. In this \textsc{Pythia}~8 model, we evaluate the gap survival probabilities using the multiparton interaction (MPI) framework.

\section{Hard diffraction and photoproduction in \textsc{Pythia~8}}

The event generation for hard diffraction in \textsc{Pythia} begins as with any other event by sampling a hard process according to the partonic cross section and the (inclusive) PDFs for the given beam types. Then, using the ratio between the diffractive and the inclusive PDF at the sampled $x$ and $Q^2$, a decision is made whether the process is of diffractive origin which would correspond to the usual factorization approach. The key step is to check whether there has been any MPIs between the beam particles and to reject such events where the rapidity gap would thus not survive and the target hadron would break up. It is, however, allowed to have further MPIs between the Pomeron mediating the diffractive scattering and the beam particle since these would not add any particles into the rapidity gap but only affect the multiplicity of the diffractive system.

The diffractive PDFs are typically derived in a QCD analysis applying diffractive DIS data in electron-proton colliders. Since the highly virtual intermediate photon does not have any internal structure but acts as a point-like particle, no violation of factorization is expected. However, in the photoproduction regime, where the intermediate photons are quasi-real, the photon may fluctuate into a hadronic state. In this case the photon become resolved and the constituents act as initiators for the hard scattering. Furthermore, a resolved photon interacting with an a hadron may also give rise to MPIs. In the hard diffraction model in \textsc{Pythia}, factorization breaking effects would thus be expected also in photoproduction. Generally the effects would, however, be less pronounced than in proton-proton collisions due to the pre\-sence of the unsuppressed direct-photon contribution.

\section{Results}

We compare the \textsc{Pythia~8} results to different HERA photoproduction data for dijet production and provide predictions for the electron-ion collider (EIC) and ultraperipheral collisions (UPCs) at the LHC. The impact of the rapidity-gap survival probability is quantified in each case by comparing the results with and without the MPI rejection. The results are based on Ref.~\cite{Helenius:2019gbd}.

\subsection{Comparisons to HERA data}

\begin{figure}[h]
\centering
\includegraphics[width=0.49\textwidth]{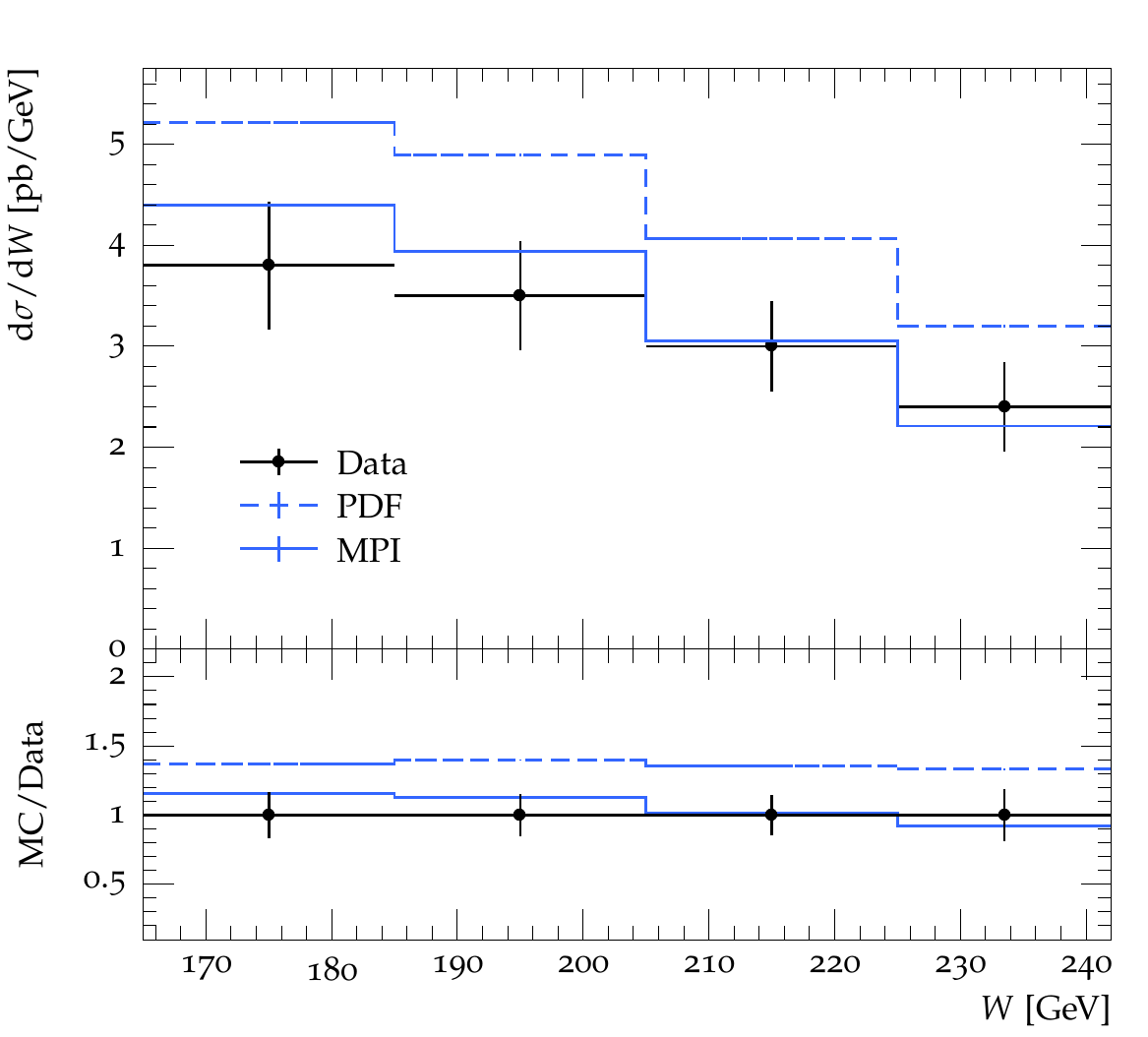}
\includegraphics[width=0.49\textwidth]{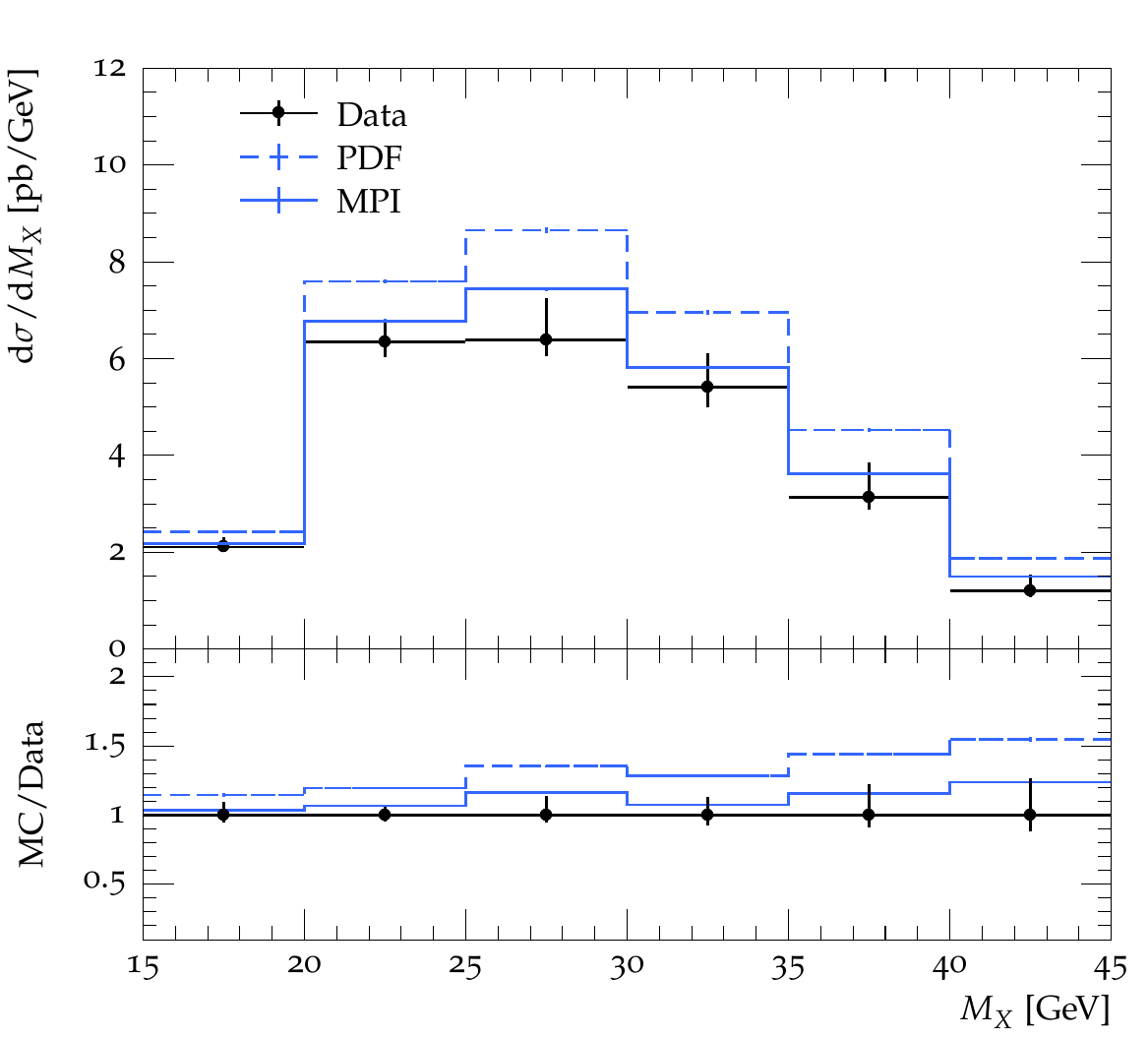}
\caption{Differential cross sections in terms of invariant mass with (solid) and without (dashed) the MPI rejection compared to H1 (left) and ZEUS (right) measurements.}
\label{fig:diffdijetWM}
\end{figure}

\begin{figure}[h]
\centering
\includegraphics[width=0.49\textwidth]{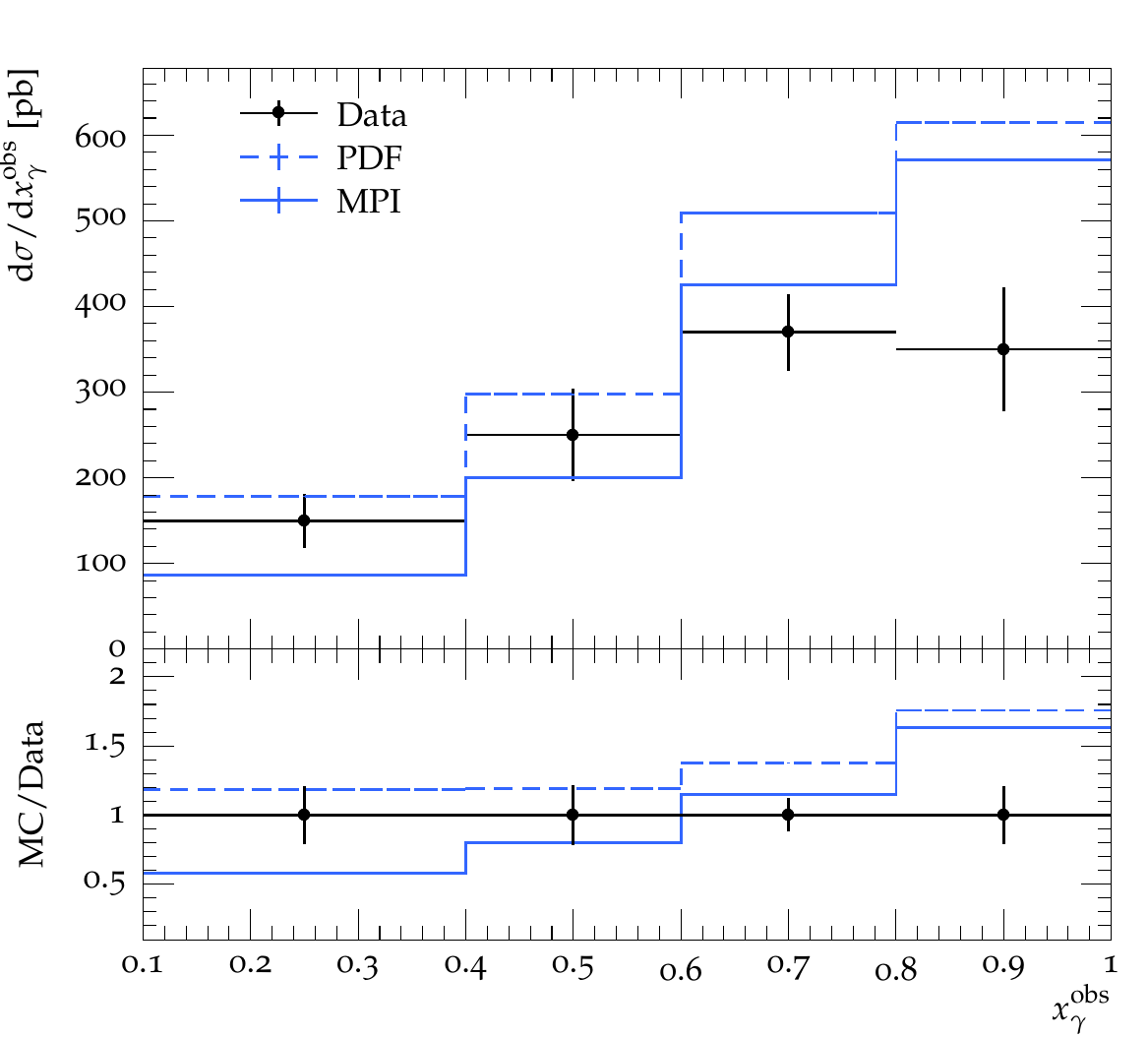}
\includegraphics[width=0.49\textwidth]{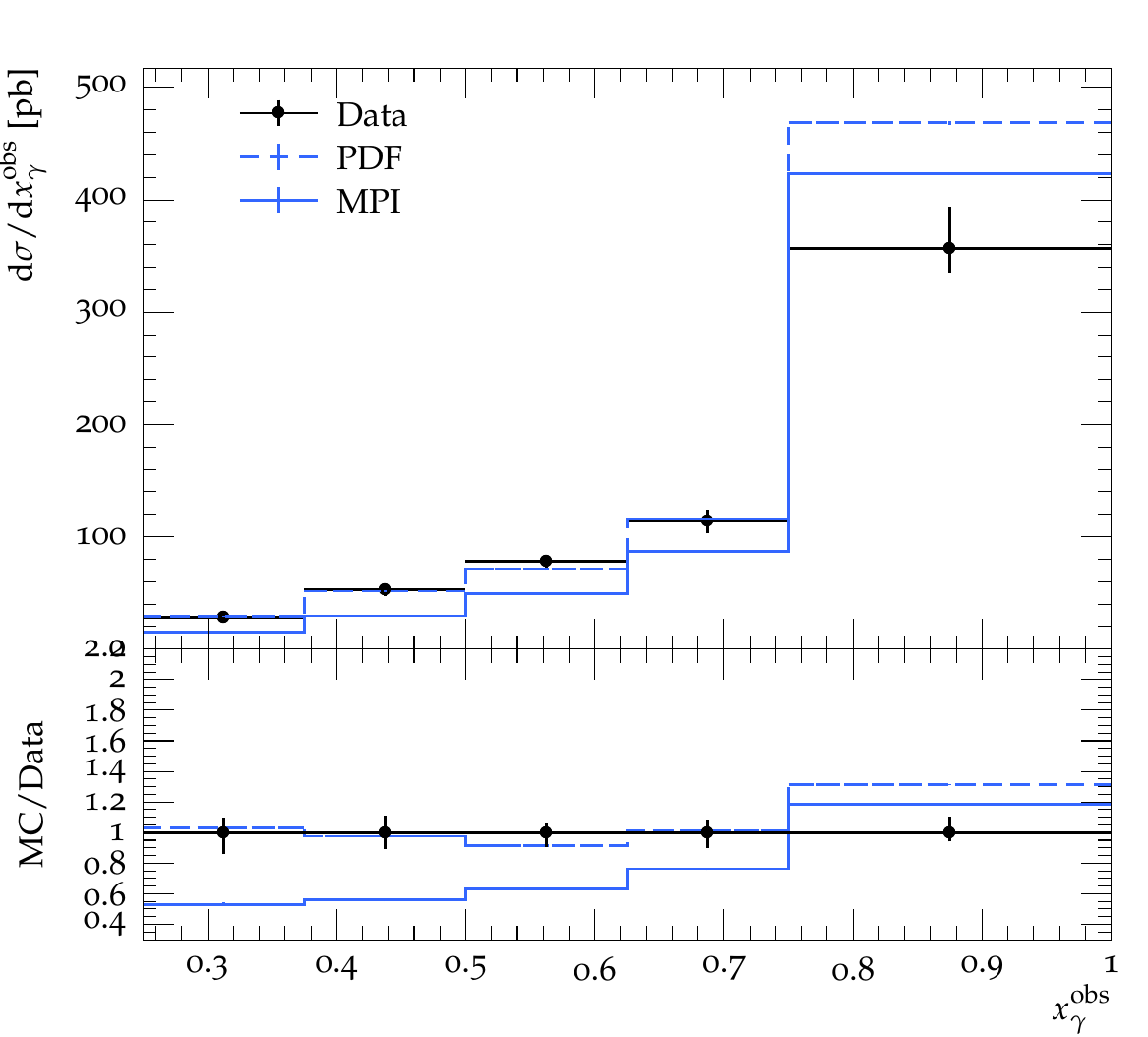}
\caption{Differential cross sections in terms of $x_{\gamma}^{\mathrm{obs}}$ with (solid) and without (dashed) the MPI rejection compared to H1 (left) and ZEUS (right) measurements.}
\label{fig:diffdijetX}
\end{figure}

We consider two different HERA analyses for photoproduction of diffractive dijets, one from H1 \cite{H1:2007jtx} and one from ZEUS \cite{ZEUS:2007uvk}. The comparisons are done by applying the prepared and published Rivet analyses \cite{Bierlich:2019rhm} for these data that can be used for further studies for other measured distributions. Here we focus on two observables that demonstrate the main features of the model, $x_{\gamma}^{\mathrm{obs}}$ which estimates the momentum fraction of the partons inside the photon, and the invariant mass $W$ ($M_X$) of the photon-proton (photon-Pomeron) system for the H1 (ZEUS) data shown in figures \ref{fig:diffdijetWM} and \ref{fig:diffdijetX}. The H1 data has been derived with a lower cut on jet-$E_{\mathrm{T}}$ of $5.0~\text{GeV}$ $(4.0~\text{GeV})$ for the (sub-)leading jet whereas in the ZEUS data the corresponding limits are $7.5~\text{GeV}$ and $6.5~\text{GeV}$. In both cases the suppression from the MPI rejection is stronger at larger invariant masses and at lower $x_{\gamma}^{\mathrm{obs}}$ which is a result of having more energy remaining for the MPIs after the jets have been created. The lower jet-$E_{\mathrm{T}}$ cuts in the H1 analysis result as stronger suppression when the MPI rejection is applied compared to the ZEUS cuts. 

\subsection{Predictions for EIC}

\begin{figure}[h]
\centering
\includegraphics[width=0.49\textwidth]{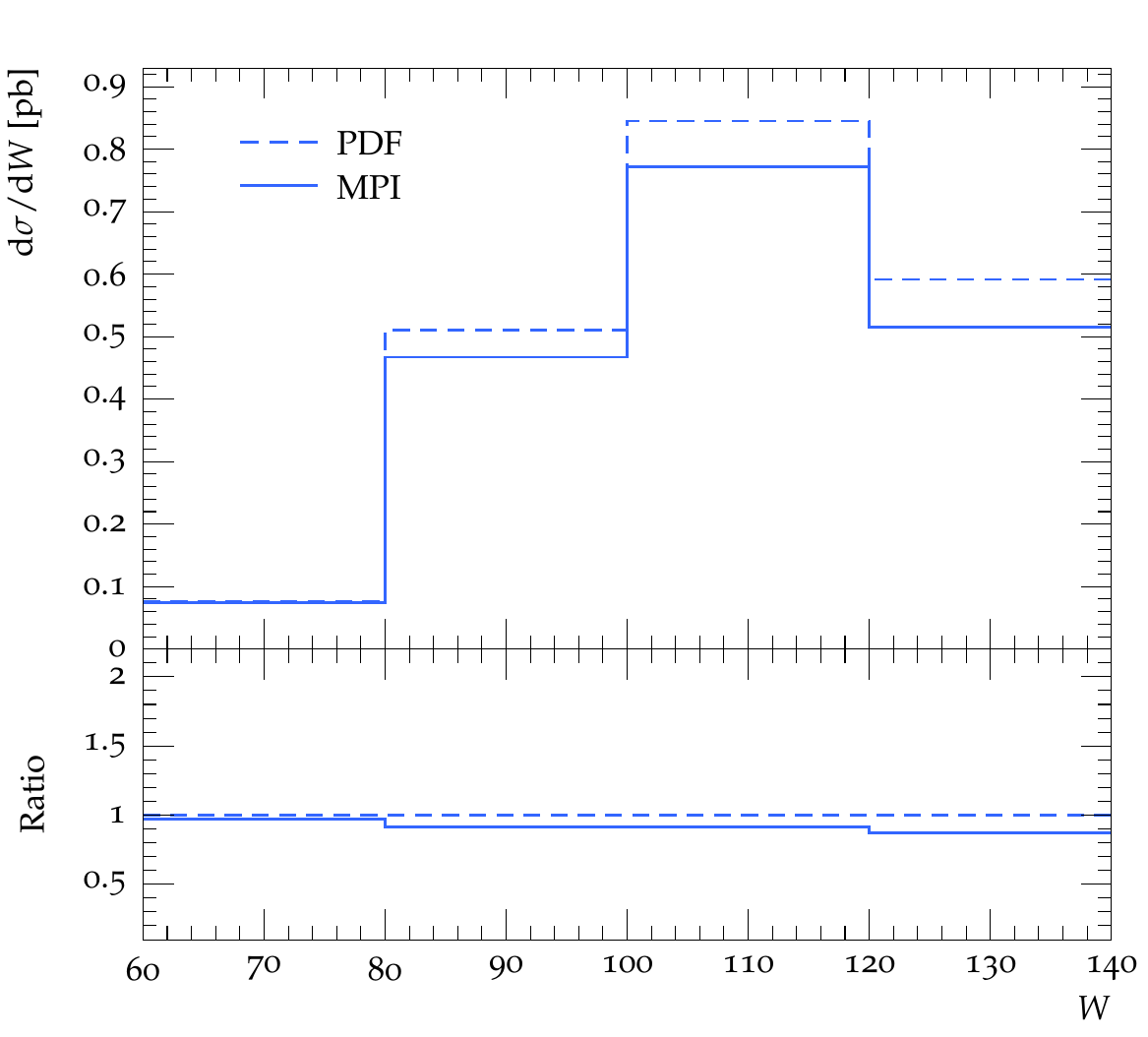}
\includegraphics[width=0.49\textwidth]{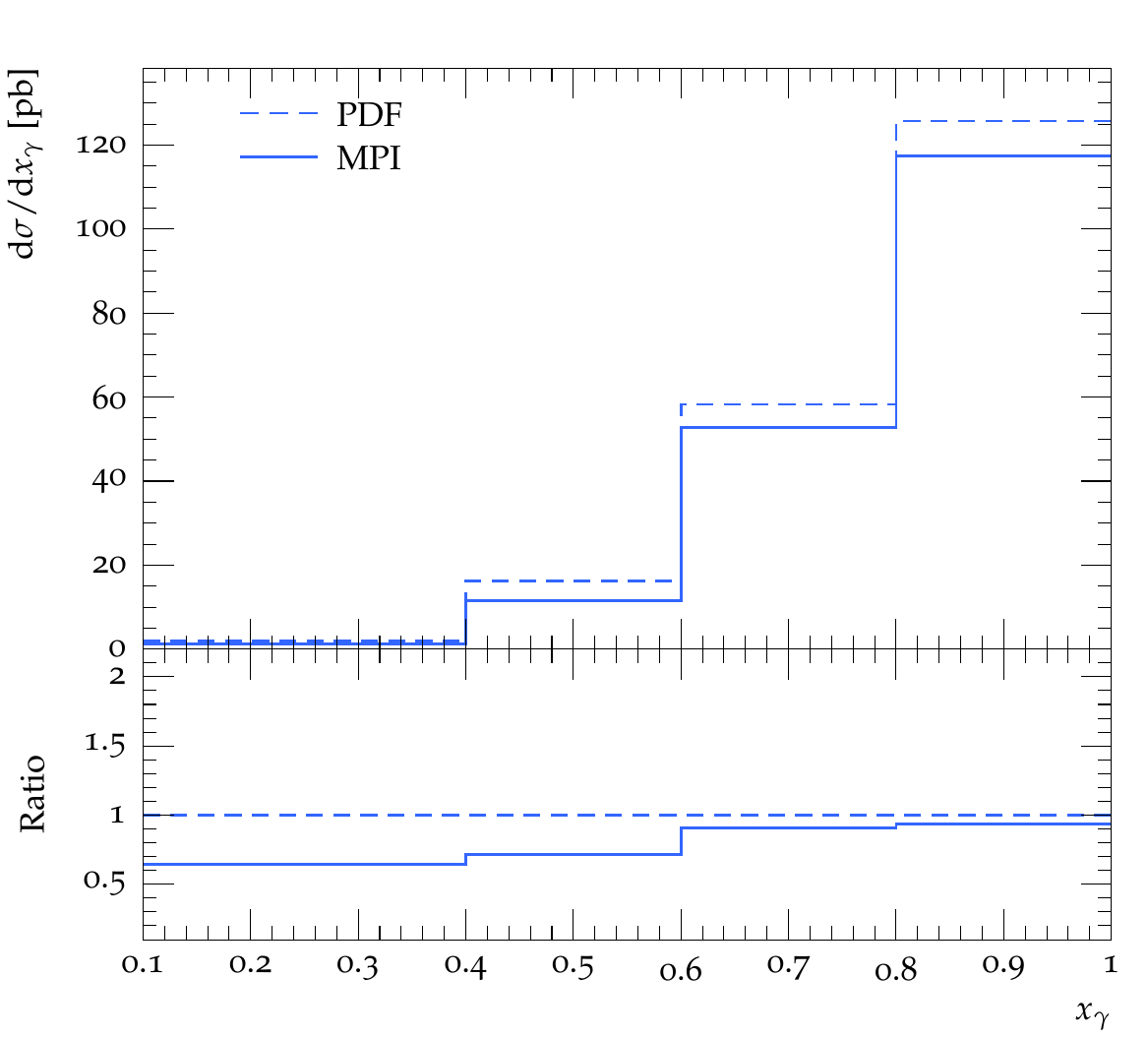}
\caption{The differential cross sections in terms of $W$ (left) and $x_{\gamma}^{\mathrm{obs}}$ (right) corresponding to EIC kinematics at $\sqrt{s} = 140~\text{GeV}$ with (solid) and without (dashed) the MPI rejection.}
\label{fig:diffdijetEIC}
\end{figure}

To estimate the expected factorization-breaking effects at the EIC kinematics, we have repeated the H1 analysis at $\sqrt{s} = 141~\text{GeV}$. The results for $W$ and $x_{\gamma}^{\mathrm{obs}}$ are shown in figure~\ref{fig:diffdijetEIC}. We find that due to the lower collision energy compared to HERA, the predicted suppression from the rapidity gap survival model is fairly modest apart from the lowest $x_{\gamma}^{\mathrm{obs}}$ where the cross section is small with the applied kinematical cuts. 

\subsection{Predictions for UPCs at the LHC}

\begin{figure}[h]
\centering
\includegraphics[width=0.49\textwidth]{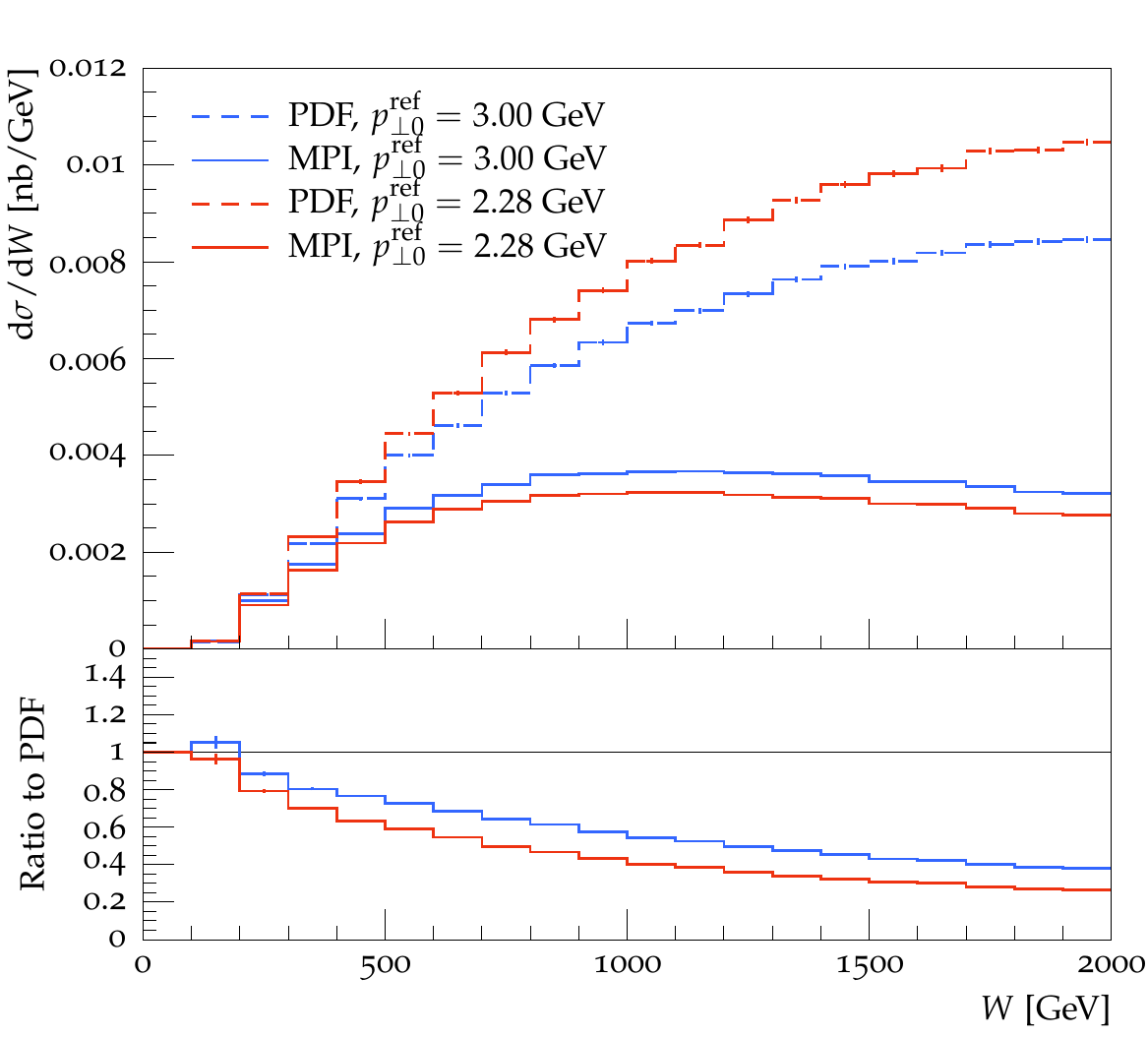}
\includegraphics[width=0.49\textwidth]{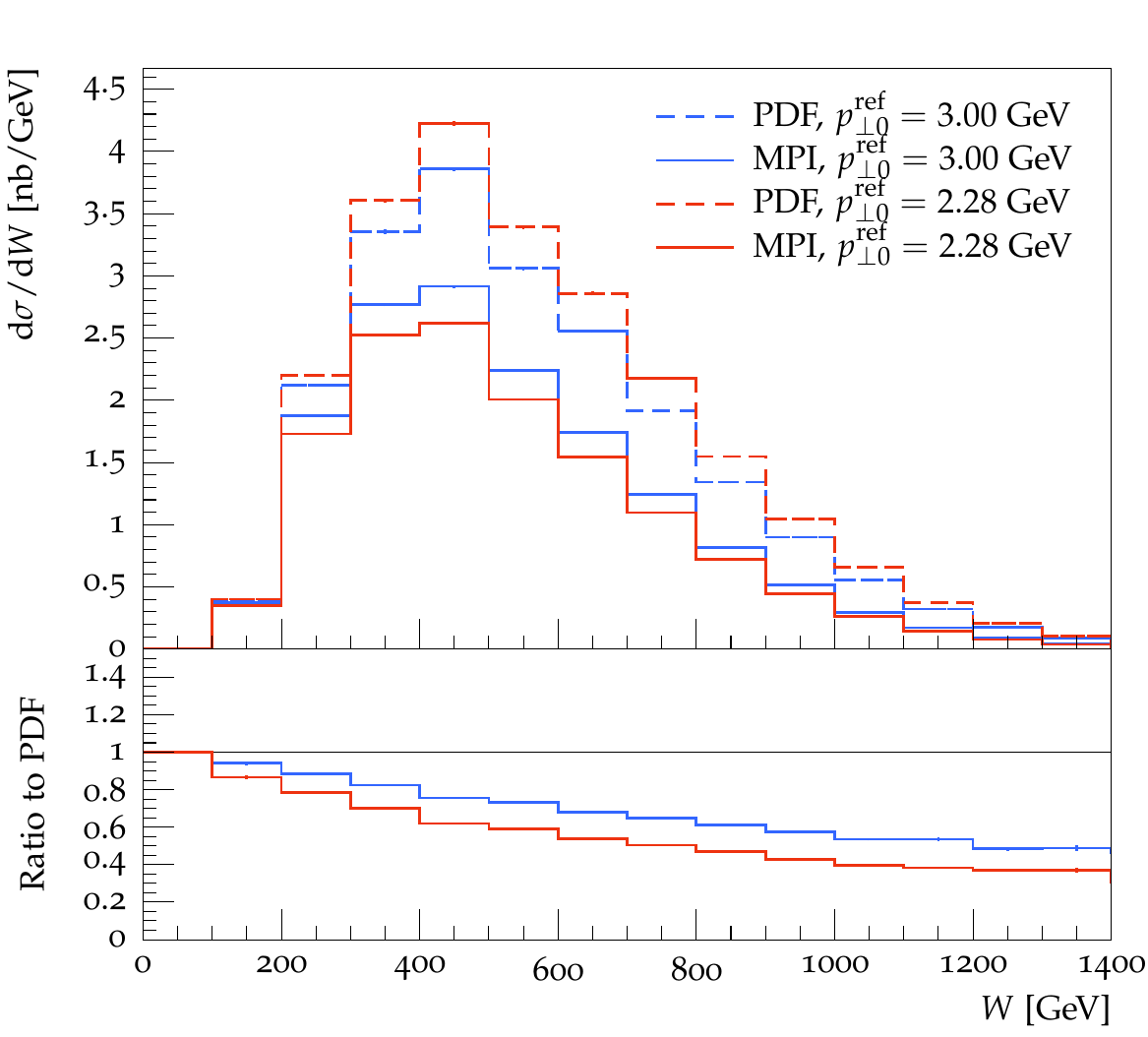}
\caption{The differential cross sections in terms of $W$ corresponding to UPCs in proton-proton (left) and proton-lead (right) collisions at the LHC with (solid) and without (dashed) the MPI rejection using the default (red) and HERA-tuned (blue) MPI parameters.}
\label{fig:diffdijetUPC}
\end{figure}

The factorization breaking effects can be studied also in UPCs where the incoming protons and nuclei do not interact directly but a photon emitted by the other beam particle collides with a particle from the another beam. Since the virtuality of the photons emitted by protons and nuclei is low, these collisions can be treated like the photoproduction in electron-proton collisions by only replacing the photon fluxes with an appropriate one. The cross sections corresponding to proton-proton and proton-lead collisions at the LHC are shown in figure \ref{fig:diffdijetUPC}. Typically the photon fluxes are somewhat softer in case of proton and nuclear beams but as the collision energy is significantly larger at the LHC ($\sqrt{s} = 5~\text{TeV}$ for pPb, $\sqrt{s} = 13~\text{TeV}$ for pp), higher values of $W$ can be reached in these UPCs than what was provided by HERA. Thus there is also more room for MPIs in these collisions and the dynamical rapidity gap survival model predicts an increased amount of suppression, especially for the proton-proton case. To estimate the underlying theoretical uncertainty in the MPI model for photon-proton collisions, the $p_{\mathrm{T,0}}^{\mathrm{ref}}$ parameter has been varied between the default proton-proton case and a value tuned to inclusive charged-particle data from HERA.

\section{Conclusion}
The dynamical rapidity gap survival model implemented for proton-proton and photon-proton collisions within the \textsc{Pythia~8} event generator provides a natural explanation for the observed fac\-to\-rization breaking effects in different collision systems and energies. Here we have studied these effects in different photon-initiated collisions and found support from the HERA data. As there is a rather strong dependence on the considered kinematics, the model could be further tested by comparing to measurements at lower collision energies (EIC) where the expected effects are mild and at higher energies (UPCs at the LHC) where the MPI suppression is expected to be stronger.


\paragraph{Funding information}
This work was supported by the Academy of Finland, Project nr. 331545.

\bibliography{helenius_DIS2021.bib}




\nolinenumbers

\end{document}